\documentclass[%
 reprint,superscriptaddress,
amsmath,
 aps, prl,twocolumn
]{revtex4-1}

\usepackage{graphicx}
\usepackage{dcolumn}
\usepackage{bm}
\usepackage{color}
\usepackage{ulem}
\newcommand{\Dai}{\textcolor{black}}
\usepackage{comment}

\DeclareUnicodeCharacter{2212}{-}

\begin{document}

\preprint{APS/123-QED}

\title{Magnetic field effect on topological spin excitations in CrI$_3$}

\author{Lebing Chen}
\affiliation{Department of Physics and Astronomy,
Rice University, Houston, Texas 77005, USA}
\author{Jae-Ho Chung}
\email{jaehc@korea.ac.kr}
\affiliation{Department of Physics, Korea University, Seoul 02841, Korea}
\author{Matthew B. Stone}
\author{Alexander I. Kolesnikov}
\author{Barry Winn}
\author{V. Ovidiu Garlea}
\author{Douglas L. Abernathy}
\affiliation{Neutron Scattering Division, Oak Ridge National Laboratory, Oak Ridge, Tennessee 37831, USA}
\author{Bin Gao}
\affiliation{Department of Physics and Astronomy,
Rice University, Houston, Texas 77005, USA}
\author{Mathias Augustin}
\affiliation{Institute for Condensed Matter Physics and Complex Systems,  School of Physics and Astronomy,
The University of Edinburgh, EH9 3FD, UK}
\author{Elton J. G. Santos}
\email{esantos@exseed.ed.ac.uk}
\affiliation{Institute for Condensed Matter Physics and Complex Systems, School of Physics and Astronomy,
The University of Edinburgh, EH9 3FD, UK}
\affiliation{Higgs Centre for Theoretical Physics, The University of Edinburgh, EH9 3FD, UK}
\author{Pengcheng Dai}
\email{pdai@rice.edu}
\affiliation{Department of Physics and Astronomy,
Rice University, Houston, Texas 77005, USA}

\date{Submitted for review on Feb. 18, 2021, Uploaded on ArXiv on \today}

\begin{abstract}
The search for topological spin excitations in recently discovered two-dimensional (2D) van der Waals (vdW) magnetic 
materials is important because of their potential applications in dissipationless spintronics. In the 2D vdW ferromagnetic (FM) honeycomb lattice 
CrI$_3$ ($T_C=61$ K), acoustic and optical
spin waves were found to be separated by a gap at the Dirac points. The presence of such a gap is a signature of topological spin excitations
\Dai{if} it arises from the next nearest neighbor (NNN) Dzyaloshinskii-Moriya (DM) or bond-angle dependent Kitaev interactions within the Cr honeycomb lattice. Alternatively, the gap is suggested to arise from an electron correlation effect not associated with topological spin excitations. 
Here we use inelastic 
neutron scattering to \Dai{conclusively} demonstrate that the Kitaev interactions \Dai{and electron correlation effects} 
cannot describe \Dai{spin waves, Dirac gap} and their in-plane magnetic field \Dai{dependence}. 
\Dai{Our results support the DM interactions being the microscopic origin of the observed 
Dirac gap}. Moreover, we find that the nearest neighbor (NN) magnetic exchange interactions along the $c$ axis 
are antiferromagnetic (AF) and the NNN interactions are FM. Therefore, our results 
unveil the origin of the observed $c$ axis AF order in thin layers of CrI$_3$, firmly determine the microscopic 
spin interactions in bulk CrI$_3$, and provide a new understanding of topology-driven spin
excitations in 2D vdW magnets. 
\end{abstract}

\maketitle

\section{Introduction}

\begin{figure}[t]
\centering
\includegraphics[scale=.38]{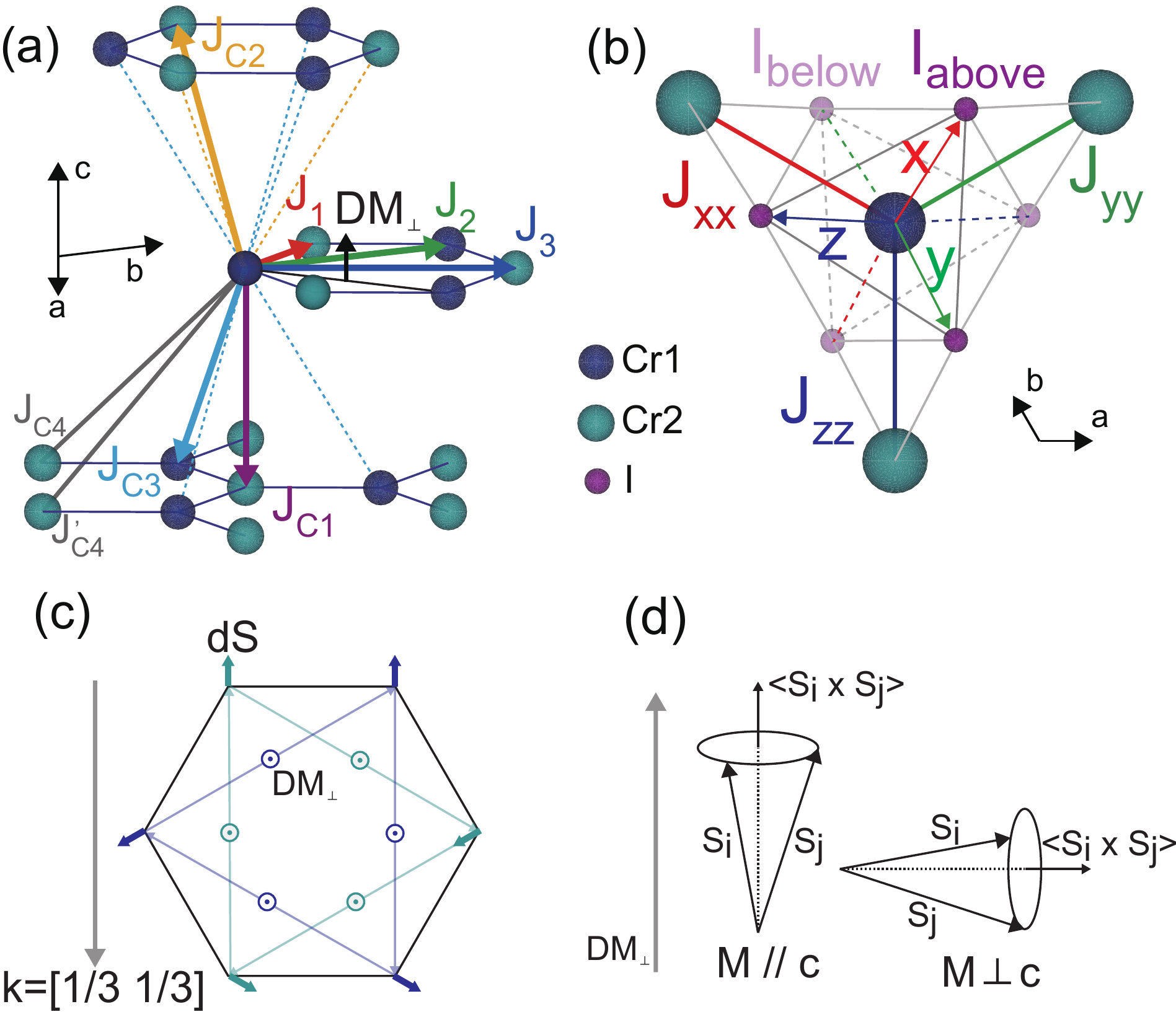}
\caption{\label{fig1}The crystal structure of CrI$_3$. (a) The CrI$_3$ rhombohedral lattice showing only Cr atoms, with Cr$^{3+}$ spins along the $c$ axis. 
\Dai{Cr1(blue) and Cr2(cyan) spheres indicate Cr atoms in different triangular sublattice.
The colored bonds indicate in-plane and interlayer magnetic exchange interactions. The cyan/yellow dashed lines show the 
3 $J_{c2}$'s and 6 $J_{c3}$'s around one Cr atom.} (b) The Kitaev interaction in the local coordinates of CrI$_3$. 
\Dai{The $J_{xx}$, $J_{yy}$, $J_{zz}$ bond is between the NNs, and the 
$\{x,y,z\}$ direction is parallel to the Cr-I bond as shown with arrows.} (c) The DM interactions in CrI$_3$ with a top view of the Cr$^{3+}$ hexagon at the Dirac wave vector. The cyan and blue colors distinguish two triangular sublattices. (d) Interactions between DM and spins. Only when spins have components along the $c$ axis, the DM term can give non-zero contribution to the total Hamiltonian. 
}
\end{figure}

The discovery of robust two-dimensional (2D) ferromagnetic (FM) long range order in monolayer van der Waals (vdW) magnets \cite{CrI3_nature,CGT_nature,CrBr3_2DFM} is important because these materials can 
provide a new platform to study fundamental physics 
without the influence of a substrate, and be potentially used to develop new spintronic devices \cite{burch2018,Gibertini2019}.
One prominent group of these materials are the chromium trihalides, CrX$_3$ (X = Br, I) or CrXTe$_3$ (X = Ge, Si), where Cr$^{3+}$ ($3d^3$, $S=3/2$) ions form 2D honeycomb lattices [Fig. 1(a)] \cite{McGuire_CrI3_2014,Mcgurie2017}. Within a single honeycomb layer, Cr$^{3+}$ ions interact with each other ferromagnetically via the nearly 90 degree 
Cr-X-Cr superexchange paths [Fig. 1(b)] \cite{G-C_rule}. Although the $3d$ electrons 
of Cr$^{3+}$ do not provide large spin-orbit coupling (SOC), the heavier ligand atoms such as iodine may serve as a source of significant SOC.  This not only provides the thermal stability observed in vdW layered
materials but also enriches the physics of magnetism in the
2D limit \cite{aniso_exchange,itinerantfermion,M-W_theorem,Kartsev20,Augustin21,Wahab20}.  \Dai{Indeed, it is proposed that 
the Kitaev interaction \cite{Kitaev}, known to be important for effective $S=1/2$ honeycomb lattice magnets near a Kitaev
quantum spin liquid \cite{Jackeli09,Takagi2019}, may occur in $S=3/2$ CrI$_3$ 
across the nearest bond with bond-dependent anisotropic Ising-like exchange [Fig. \ref{fig1}(b)].  This would be critical for the magnetic stability of monolayer CrI$_3$ and spin dynamics in bulk CrI$_3$ \cite{Xu2018,Kitaev_ILee2020,Stavropoulos2021,Kitaev_Josh2018,Kitaev_Aguiliera2020}.} 
\Dai{Furthermore, }spin waves (magnons) from honeycomb ferromagnets can be topological by opening a gap at the Dirac points via time-reversal symmetry breaking (TRSB) \cite{Owerre,SKKim}. As a magnetic analog of electronic dispersion in graphene \cite{Neto},  
spin wave spectra of honeycomb ferromagnets have Dirac points at the Brillouin zone boundaries where dispersions of acoustic and optical spin waves meet  
and produce Dirac cones. If the system has TRSB arising from a large SOC,
one would expect to observe an energy gap at the Dirac point of the bulk magnon  bands \cite{Owerre}, analogous to the 
SOC induced gap at the Dirac point in the electronic 
dispersion of graphene \cite{Haldane}. This in turn would allow 
the realization of mass-less 
topological spin excitations propagating without dissipation \cite{Chumak2015,Chernyshev2016,Wang2017}.

Experimentally, a spin gap was indeed observed at the Dirac point in the spin wave spectra of the honeycomb lattice FM CrI$_3$ \cite{LBChen}. \Dai{Three}  possible scenarios have been proposed to understand the observed spin gap.  The first corresponds to the Dzyaloshinskii-Moriya (DM) interaction that occurs 
on the bonds without inversion symmetry [Figs. \ref{fig1}(a,c,d)] \cite{DM1,DM2}. The second scenario is the Kitaev interaction that also breaks the time reversal symmetry and can inhabit nontrivial topological edge modes \cite{Kitaev_ILee2020,LBChen2}. Finally, the observed Dirac spin gap is suggested to arise from electron correlations that must be treated explicitly to understand the spin dynamics in CrI$_3$ and the broad family
of 2D vdW magnetic materials \cite{Ke2021}. In this case, spin excitations in CrI$_3$ would not be topological.

Another intriguing property of CrI$_3$ is its weak structural and magnetic coupling along the $c$ axis. In the low-temperature FM phase, bulk CrI$_3$ \Dai{is assumed to} have  rhombohedral lattice structure with space group $R\bar{3}$ \cite{McGuire_CrI3_2014}. 
On warming across $T_C$, the FM order in CrI$_3$ 
disappears in a weakly first order phase transition coupled with a small $c$ axis lattice \Dai{parameter change}. 
Upon further warming to 90-200 K, CrI$_3$ undergoes a first order phase transition from rhombohedral to monoclinic 
structure with $C/2m$ space group, basically shifting the stacking of
the CrI$_3$ layers \cite{McGuire_CrI3_2014}. From comparisons to spin wave dispersions, the nearest neighbor (NN) $c$ axis magnetic exchange coupling was deduced to be FM with $J_{c1}\approx 0.59$ meV [Fig. 1(a)] \cite{LBChen}. However, transport, Raman scattering, scanning magnetic circular dichroism microscopy, and tunneling measurements as a function
of film thickness \cite{CrI3_nature,XXZhang2020,JCenker2021}, pressure \cite{Pressure_CrI3, Pressure2}, and applied magnetic field \cite{MFCrI3} suggest $A$-type antiferromagnetic (AF) structure
associated with monoclinic structure 
present in bilayer and a few top layers of bulk CrI$_3$.  
In particular, a magnetic field a few Tesla along the $c$ axis was found to modify the crystal lattice symmetry of CrI$_3$, thus 
suggesting a strong spin-lattice coupling \cite{MFCrI3}. 
Therefore, 
it is important to determine if the NN interlayer exchange coupling is indeed FM, and what determines the overall 
FM interlayer coupling in CrI$_3$ bulk with rhombohedral lattice structure.

\begin{figure}[t]
\centering
\includegraphics[scale=.40]{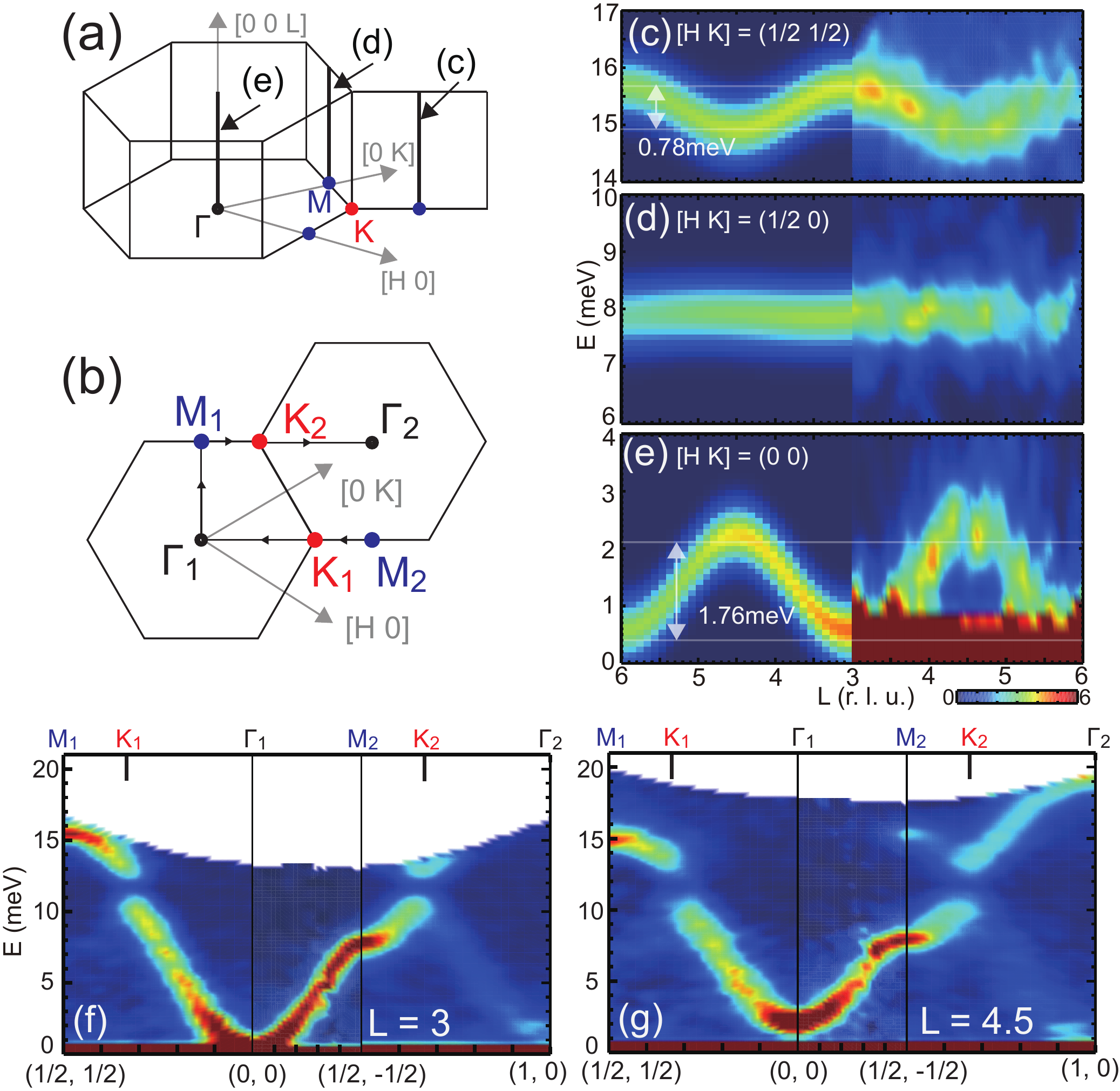}
\caption{\label{fig2}\Dai{Spin wave spectra of CrI$_3$. (a) The hexagonal reciprocal lattice of CrI$_3$. Gray arrows show reciprocal lattice vectors, and high-symmetry ($\Gamma,K,M$) points are specified in blue ($M$), red ($K$), and black ($\Gamma$) dots, respectively. 
The bold black lines specify the scan direction in (c-e). (b) Projection of the hexagonal reciprocal lattice in the $[H,K]$ plane. The arrows indicate the scan path of the spectra shown in (f, g), Fig. 3(a-c), and Fig. 4(a,b). (c-e) The spin wave dispersion along the $L$ direction 
at different $[H, K]$ positions specified in (a), showing different bandwidths at different $[H,K]$ points. The left and right panels are calculation and data, respectively. (f, g) Spin wave dispersion at different $L$ points. (f) shows $L$ integration range $[2.5,3.5]$ near the $[0,0,L]$ band bottom,  
while (g) shows $L$ integration range $[4,5]$ near the band top.}
}
\end{figure}

In this work, we use high-resolution inelastic neutron scattering to study spin waves
of CrI$_3$ and their magnetic field dependence. By reducing the mosaic of co-aligned
single crystals of CrI$_3$ from 
earlier work \cite{LBChen}, we were able to precisely measure the 
magnitude of the spin gap at the Dirac points and \Dai{the entire spin wave spectra}. 
In addition, \Dai{we determine the effect of an in-plane magnetic field on spin waves and Dirac spin gap in CrI$_3$. 
By comparing the experimental observations with expectations from Heisenberg-DM and Heisenberg-Kitaev Hamiltonian, 
and the effect of electron correlations, we conclude that spin waves and Dirac spin gap 
in CrI$_3$ cannot be described by the Heisenberg-Kitaev Hamiltonian and electron correlation effects.  Instead, the data 
are approximately consistent with the Heisenberg-DM Hamiltonian with considering both the $c$ axis and in-plane DM interactions.}  
  Our results therefore clarify the microscopic spin interactions in CrI$_3$ and provide a new understanding of topology-driven spin
excitations in 2D vdW magnets.

\section{Results}

Single crystalline CrI$_3$ samples were grown using the chemical vapor transport method as described in Ref. \cite{McGuire_CrI3_2014}. Our inelastic neutron scattering experiments were carried out on either fully co-aligned ($\sim$ 0.42 g) or $c$ axis aligned ($\sim$ 1 g) crystals on the SEQUOIA \cite{SEQUOIA}, HYSPEC \cite{HYSPEC}, and ARCS \cite{ARCS}
spectrometers at Spallation Neutron Source, Oak Ridge National Laboratory. 
Consistent with Ref. \cite{LBChen}, we use honeycomb lattice with 
in-plane Cr-Cr distance of $\sim$3.96 \AA\ and $c$ axis layer spacing of 6.62 \AA\ in the low temperature rhombohedral structure to describe CrI$_3$. The momentum transfer $\textbf{Q}=H\textbf{a}^\ast+K\textbf{b}^\ast+L\textbf{c}^\ast$ is denoted as $(H,K,L)$ in reciprocal lattice units (r.l.u.) with marked 
high symmetry points [Figs. \ref{fig2}(a,b)]. All measurements were carried out with the $c$ axis of the sample in the horizontal scattering plane, and the applied magnetic fields vertical, i.e., in the $ab$ plane of CrI$_3$ \Dai{[Figs. \ref{fig1}(a,c,d)]}.

\begin{figure}[t]
\centering
\includegraphics[scale=.40]{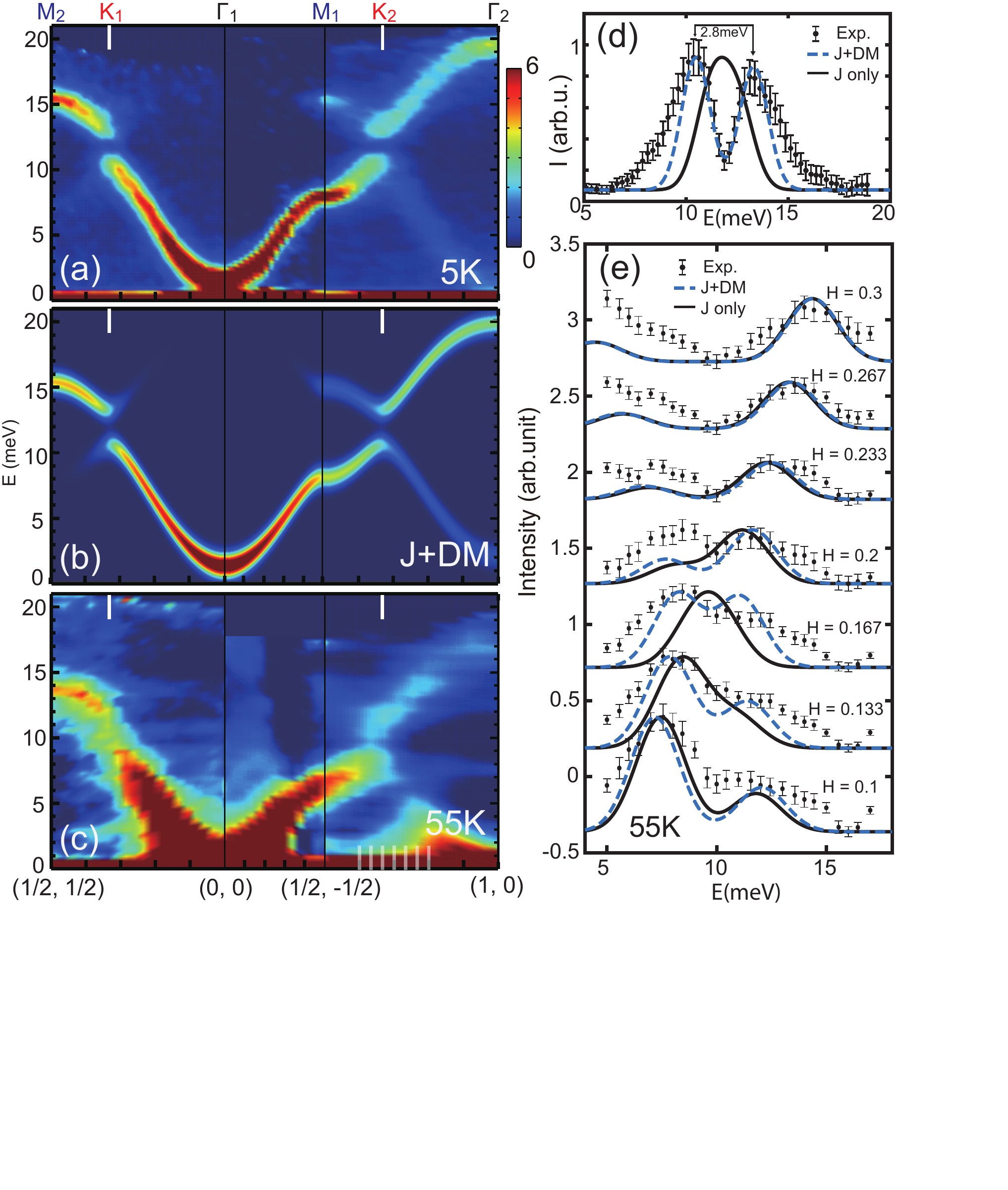}
\caption{\label{fig3}The Heisenberg-DM model fit of CrI$_3$ $E$-${\bf Q}$ spin wave spectrum.  
High-symmetry points are labeled. (a) Experimental data at 5 K. (b) Heisenberg-DM model fits convoluted with instrumental resolution; 
(c) Experimental data at 55 K; The $L$ integration range in (a-c) is $[-6, 6]$. (d) Energy cuts at the $K_2$ points from (a) and (b), 
showing a Dirac gap of $\sim$2.8 meV; (e) Energy cuts at different wave vectors of the 55 K data as shown the shaded white lines in (c). Blue dashed lines are Heisenberg-DM fits, and black solid lines show the same model with zero DM term in (e,f).
}
\end{figure}

We begin by describing the zero field high-resolution spin wave data of CrI$_3$ obtained on SEQUIOA [\Dai{Figs. \ref{fig2} and \ref{fig3}}]. Figure \ref{fig3}(a) shows the energy-momentum 
($E$-${\bf Q}$) dependent spin wave spectra along the high-symmetry directions 
in reciprocal space as depicted in Fig. \ref{fig2}(a). 
These in-plane spin wave spectra were obtained by integrating dispersive spin waves
along the $c$ axis over $-6 \leq L \leq 6$. The overall momentum dependence of the spin wave energies are consistent with previous work \cite{LBChen}, revealing two spin wave modes characteristic of the honeycomb ferromagnets. The lower and upper modes account for the acoustic and optical vibrations, respectively, of the two sublattice spins. These two spin wave modes will meet each other at the Dirac wave vectors of $Q_{K_1} = (\frac{1}{3}, \frac{1}{3})$ and $Q_{K_2} = (\frac{2}{3}, -\frac{1}{3})$ [Figs. 3(a) and 3(d)]. Inspection of Fig. 3(a) reveals clear evidence of a spin gap of $\sim$2.8 meV, which is approximately 50\% the value estimated from previous low-resolution 
data \cite{LBChen}. This is mostly due to reduced mosaicity of the co-aligned single crystals [an in-plane mosaic full-width-half-maximum (FWHM) of 8.0$^\circ$ compared with that of $\sim$17$^\circ$ in Ref. \cite{LBChen}] and improved instrumental resolution \cite{SI}. 

\begin{figure*}[t]
\centering
\includegraphics[scale=.44]{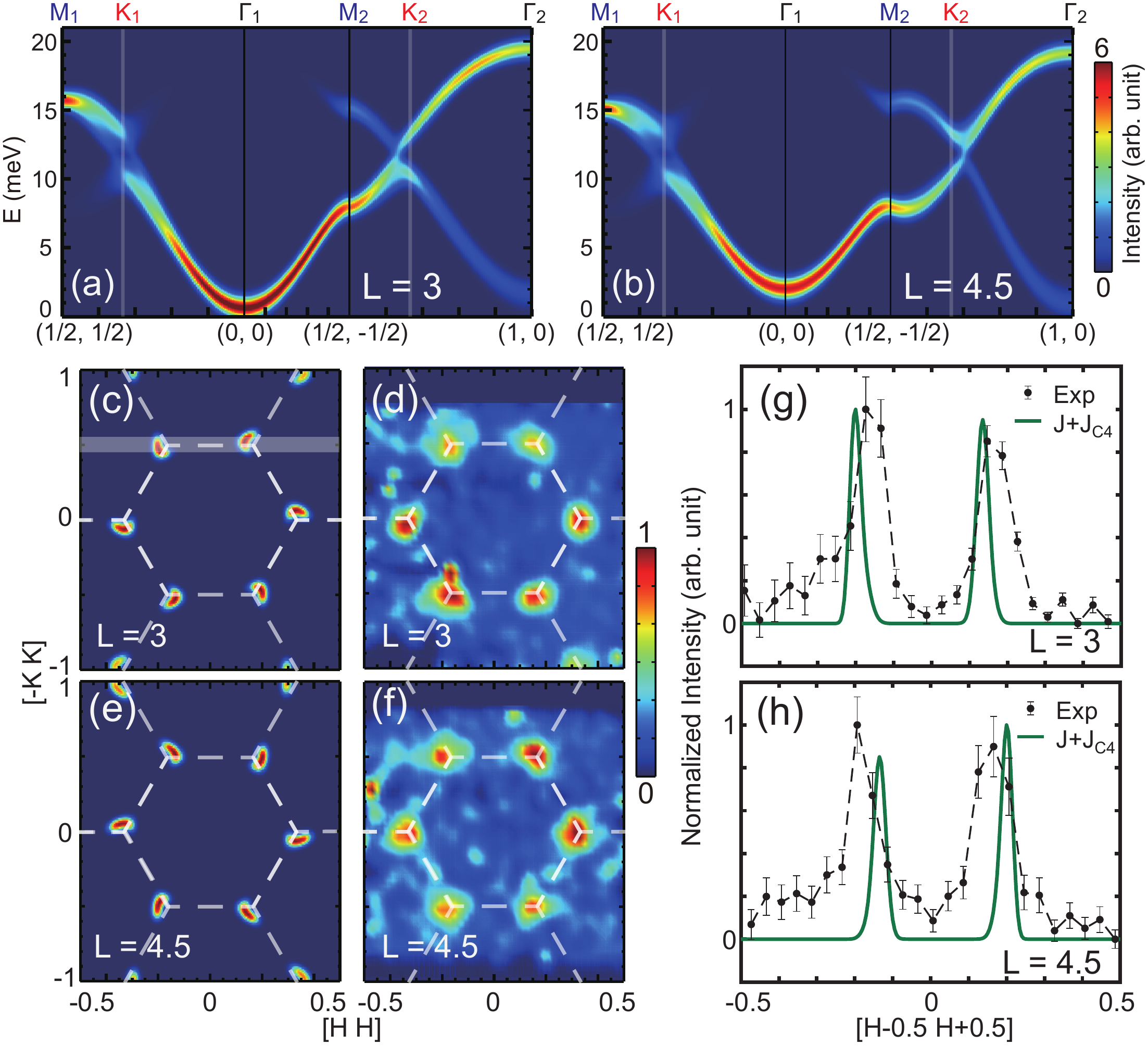}
\caption{\label{fig4}\Dai{The Heisenberg-$J_{c4}$ (spin correlation) model fit of CrI$_3$ spin wave spectra, here 
$J_{c4} = -J_{c4}^\prime = -0.193$ meV. (a, b) In-plane spin excitation spectrum with $L$ integration range $[2.5, 3.5]$ and $[4, 5]$, respectively. 
(c-f) Constant-$E$ scan at $E=12\pm 0.5$ meV with aforementioned $L$ integration range, panels (c, e) show calculation results, and panels (d, f) show experimental data. The white dashed lines are the Brillouin zone boundaries. (g, h) ${\bf Q}$-cuts along the $[H, H]$ with $[-K, K]$ integrated between $[0.45, 0.55]$ [white shaded area in (c)], black dots show the experimental data while the green solid lines 
show the Heisenberg-$J_{c4}$ model calculation.}
}
\end{figure*}

To completely determine the spin wave spectra of CrI$_3$, we show in 
Figs. \ref{fig2}(c)-\ref{fig2}(e) the $L$-dependence of spin waves
at different in-plane wave vectors. Inspection of the figures reveals  
that the modes along the $[\frac{1}{2},\frac{1}{2},L]$ and $[0,0,L]$ directions
exhibit mutually opposite $L$-dependence. Along the $[0,0,L]$ direction, the
 spin wave dispersion exhibits minimum of 0.4 meV at $L = 3n$ 
($n$ = integers) and maximum of 2.1 meV at $L = 3n+\frac{3}{2}$ [Fig. \ref{fig2}(e)]. 
In contrast, the mode along the $[\frac{1}{2},\frac{1}{2},L]$
direction peaks at $L = 3n$ and has minimum at $L = 3n+\frac{3}{2}$ [Fig. \ref{fig2}(c)],
while spin waves along the $[\frac{1}{2}, 0, L]$ 
direction are featureless [Fig. \ref{fig2}(d)].
\Dai{The overall spin wave spectra at $L=3$ and 4.5 are shown in Figs. \ref{fig2}(f) and (g),
respectively.} 
The opposite $L$ dependence between the high- and low-energy spin waves
requires finite FM inter-plane exchanges along the bonds that are tilted off the $c$ axis. 

\Dai{To understand spin wave spectra in Figs. 2 and 3, we consider a Heisenberg model with the DM interaction to account
for the observed Dirac spin gap \cite{Owerre,SKKim,LBChen}.}
The Hamiltonian of the DM interaction, $H_{DM}$, can be written as
$H_{DM}=-\sum_{i<j} \left[ {\bf A}_{ij}\cdot ({\bf S}_i\times{\bf S}_j)\right]$,
where ${\bf S}_i$ and ${\bf S}_j$ are spins at site $i$ and $j$, respectively,
and ${\bf A}_{ij}$ is the antisymmetric DM interaction between sites $i$ and $j$ [Figs. \ref{fig1}(a,c)]. 
\Dai{The combined Heisenberg-DM (J-DM) Hamiltonian is 
$H_{J-DM} = \sum_{i<j}{[J_{ij}\textbf{S}_{i}\cdot\textbf{S}_{j}+\textbf{A}_{ij}\cdot\textbf{S}_{i}\times\textbf{S}_{j}]}+\sum_{j}{D_{z}(S_{j}^{z})^{2}}$, where $J_{ij}$ is magnetic exchange coupling of the ${\bf S}_i$ and ${\bf S}_j$,
and $D_z$ is the easy-axis anisotropy along the $z (c)$ axis \cite{LBChen}. 
As shown in Fig. \ref{fig1}(a), we define the in-plane NN, the next nearest neighbor (NNN), and the third
NN interactions as $J_1$, $J_2$, and $J_3$, respectively.  The $c$ axis NN, the NNNs, and the third NN interactions are
$J_{c1}$, $J_{c2}/J_{c3}$, and $J_{c4}/J^\prime_{c4}$, respectively.
} For ideal honeycomb lattice materials
where the NNN bond breaks the inversion symmetry [Fig. 1(c)], the DM vectors can have both 
in-plane and out-of-plane components, but the former will not contribute to the topological gap opening due to the three-fold rotational symmetry of the honeycomb lattice [Fig. 1(d)].  As a result, only the DM term parallel to the $c$ axis, i.e., the NNN DM interaction, 
will contribute to the opening of a spin gap in
spin wave spectra.  Since bulk CrI$_3$ orders ferromagnetically
below a Curie temperature of $T_C\approx 61$ K with an ordered moment 
along the $c$ axis \cite{McGuire_CrI3_2014}, one can fit the spin wave spectra and Dirac gap using the finite NNN $H_{DM}$ ($\neq 0$) that may induce TRSB and topological spin excitations in the FM ordered state \cite{LBChen}. 

\begin{table}[]
    \centering
    \begin{tabular}{c|c|c|c}
    \hline
       Model & {\rm J}-{\rm DM}  & \rm{J}-${\rm J_{c4}}$ & \rm{J}-{\rm K}-$\Gamma$ \\
       \hline
        $J_1$ (meV) & -2.11 & -2.11 & -0.28 \\
        $J_2$ (meV) & -0.11 & -0.11 & -0.21 \\
        $J_3$ (meV) & 0.10 & 0.10 & 0.05 \\
        $J_{c1}$ (meV) & 0.048 & 0.048 & 0.048 \\
        $J_{c2}$($J_{c3}$) (meV) & -0.071 & -0.071  & -0.071 \\
        $J_{c4}$($-J_{c4}'$) (meV) & 0 & -0.193 & 0 \\
        $DM_{\perp}$ (meV) & 0.17 & 0 & 0 \\
        $K$ (meV) & 0 & 0 & -5.45 \\
        $D_z$ (meV) & -0.123 & -0.123 & 0 \\
        $\Gamma$ (meV) & 0 & 0 & -0.082 \\
        \hline
    \end{tabular}
    \caption{The magnetic exchange interaction strength (negative value indicates FM exchange) in the J-DM model, the electron correlation model, and the J-K-$\Gamma$ model.}
    \label{tab:my_label}
\end{table}

The left panels of Figs. \ref{fig2}(c-e)
and Fig. 3(b) are the calculated spin wave spectra with exchange parameters listed in Table I  \cite{LBChen}. 
Given the nearly flat dispersion 
along the $[\frac{1}{2},0,L]$ direction shown in Fig. \ref{fig2}(d), we chose to set $J_{c2}=J_{c3}$ for the two interplane NNN exchanges of nearly identical bond lengths [Fig. 1(a)]. The best-fit parameters reveal that the NN interlayer magnetic 
interactions are AF with strong FM
couplings along the NNN directions [Fig. \ref{fig1}(a)].  In addition, one must
include finite DM interaction $A$ to account for the observed spin gap at the
Dirac points [Figs. \ref{fig3}(a,b,d)] \cite{SI}. Figure \ref{fig3}(c) shows spin wave spectra at $T=0.9T_C=55$ K, which again shows a possible gap opening at the Dirac points.   
Figure \ref{fig3}(e) plots energy cuts at different wave vectors near the Dirac point, where the black solid lines and red dashed lines are calculations with and without the NNN DM interactions, respectively. It is clear that the excitations at $H \approx 0.167$ consist of two peaks separated by a spin gap indicating that the Dirac gap persists still at $T=0.9T_C$. Since the Dirac wave vector is along the zigzag bonds of the honeycomb lattice, the observation of a spin gap at the Dirac point indicates a  
symmetry breaking field between the two Cr sublattices within the honeycomb lattice 
[Figs. \ref{fig1}(a,c)]. While the DM vectors may be oriented either along the $c$ axis or perpendicular to it, only the 
$c$ axis component can open the Dirac gap due to the three-fold symmetry of the ideal honeycomb lattice. In addition, the magnitude of the gap is directly proportional to the $c$ axis component of the ordered spins \cite{inplaneDM_kvashnin,inplaneDM_Mook}.

\begin{figure*}[t]
\centering
\includegraphics[scale=.30]{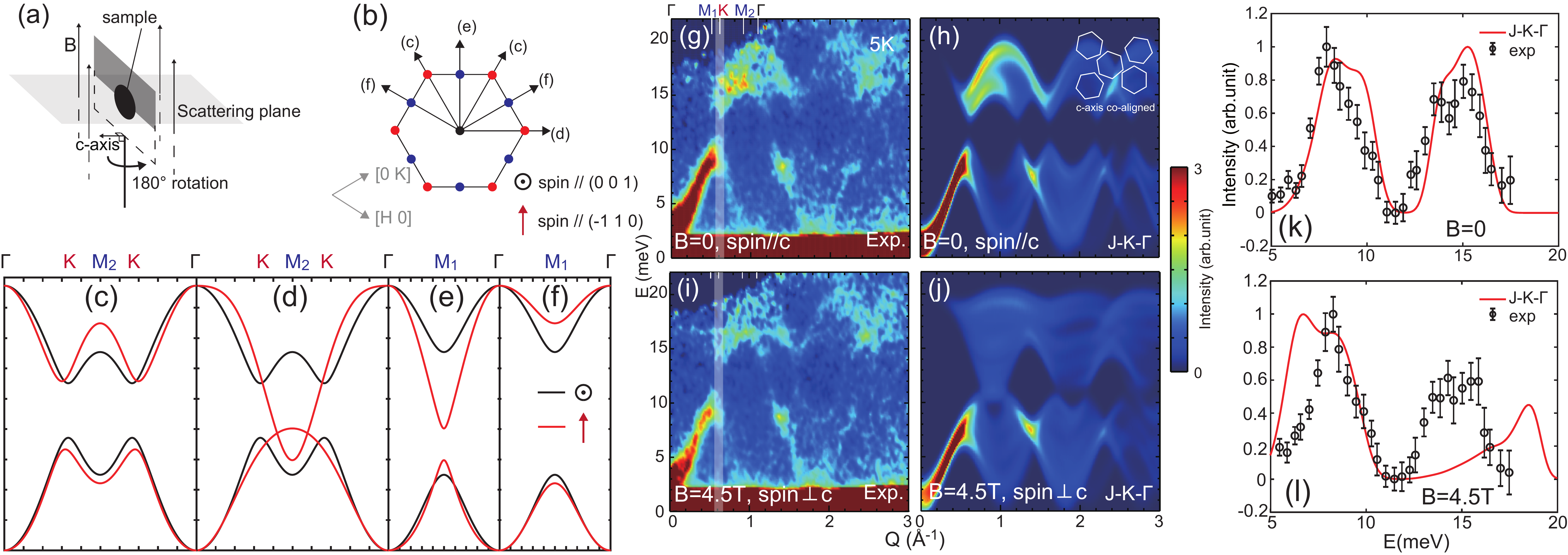}
\caption{\label{fig5}\Dai{The in-plane magnetic field effects on spin waves of 
$c$ axis aligned CrI$_3$ single crystals shown in the inset of (h), and Heisenberg-Kitaev model fit of the spectra. 
(a) The experimental setup of inelastic neutron scattering experiments, where applied field is vertical and 
$c$ axis of the crystals is in the light-shaded horizontal scattering plane. 
(b) The reciprocal lattice showing the scan direction in (c-f). The high symmetry points are shown with blue ($M$), red ($K$) and black ($\Gamma$) dots. (c-f) Spin wave Dispersions of the Heisenberg-Kitaev model with in-plane (red) and out-of-plane (black) spin orientations. 
(g, i) Spin wave $E$-\textbf{Q} spectra of CrI$_3$ at 5 K in zero and 4.5 T in-plane fields, respectively. The high-symmetry points are marked on top. Here \textbf{Q} in the unit of $\AA^{−1}$indicates the wave vector’s projection on the $[H, K]$ plane with $L= [−5,5]$ integration. 
(h, j) Calculated $E$-\textbf{Q} spectra using the Heisenberg-Kitaev Hamiltonian at zero and 4.5 T field, respectively. (k, l) Comparison of the energy cuts between experiments (black dots) and calculations (red lines) using the Heisenberg-Kitaev Hamiltonian at Dirac point in 0 T and 4.5 T, respectively. The \textbf{Q} integration range of the energy cuts is 0.55-0.66 \AA$^{−1}$ 
centered around the $K$ point (=0.608 \AA$^{−1}$), as shown in the long white shaded line in (g) and (i).}
}
\end{figure*}

An alternative scenario to understand the observed spin gap at the Dirac point is through the Kitaev interaction that occurs across the nearest bond with bond-dependent anisotropic Ising-like exchange [Fig. \ref{fig1}(b)] \cite{Kitaev}, which also breaks the time reversal symmetry and can inhabit nontrivial topological edge modes \cite{Kitaev_ILee2020}. The Kitaev interaction
Hamiltonian $H_K$ is $ H_K=\sum_{<ij>\in\lambda\mu(\nu)}[K{S}_{i}^{\nu}{S}_{j}^{\nu}+\Gamma({S}_{i}^{\lambda}{S}_{j}^{\mu}+{S}_{i}^{\nu}{S}_{j}^{\lambda})]$, 
where ($\lambda$, $\mu$, $\nu$) are any permutation of $(x, y, z)$, $K$ is the
strength of the Kitaev interaction, and $\Gamma$ is the symmetric off-diagonal anisotropy that induces a spin gap at the $\Gamma$ point \cite{Kitaev_ILee2020,LBChen2}. \Dai{The combined Heisenberg-Kitaev Hamiltonian, the so-called J-K-$\Gamma$ Hamiltonian, is 
$H_{J-K-\Gamma} = \sum_{<ij>\in\lambda\mu(\nu)}{[J_{ij}\textbf{S}_{i}\cdot\textbf{S}_{j}+K{S}_{i}^{\nu}{S}_{j}^{\nu}+\Gamma({S}_{i}^{\lambda}{S}_{j}^{\mu}+{S}_{i}^{\nu}{S}_{j}^{\lambda})]}$.} By fitting the J-K-$\Gamma$ Hamiltonian using the data shown in Fig. 3(a), we extract the exchange parameters shown in Table I, which is overall consistent with Ref. \cite{LBChen2}. When FM ordered spins are oriented along the $c$ axis \cite{McGuire_CrI3_2014}, 
the spin Hamiltonian based on the Heisenberg-Kitaev exchanges can also reproduce the observed spin waves and energy gap at the Dirac point in CrI$_3$ \cite{LBChen2}.  Therefore, one cannot determine whether the NNN DM or Kitaev model is responsible for the spin gap at Dirac points in 
the spin waves of CrI$_3$ at zero field \cite{LBChen2}.

Finally, by using calculations beyond density functional theory (DFT), 
it was suggested that the observe Dirac spin gap arises from the electron correlations not considered 
in the usual DFT theory \cite{Ke2021}. In this picture, the Dirac spin gap arises from the differences in $c$ axis
magnetic exchange pathways along the third NN $J_{c4}$ and $J^\prime_{c4}$ [Fig. 1(a), and see Fig. 3 in Ref. \cite{Ke2021}]. If this picture is correct, one would expect
Dirac nodal lines, where acoustic and optical spin wave bands cross, wind around the Dirac $K$ point along the $L$ direction \cite{Ke2021}. 
Since both $J_{c4}$ and $J^\prime_{c4}$ connect with Cr1 and do not break the CrI$_3$ sublattice symmetry, 
the electron correlation effects do not produce a true Dirac spin gap, and only 
cause the Dirac crossing to shift sideways and induce nodal winding along the $c$ axis. 
The spin wave intensity winding around the Dirac point has been observed in the insulating easy-plane 
honeycomb quantum magnet CoTiO$_3$ without
a Dirac spin gap and DM interaction, suggesting the non-trivial topology of the
Dirac magnon wavefunctions \cite{DiracMagnon,inplaneDM_McClarty,CoTiO3_Yuan,CoTiO3_Elliot}.

\begin{figure*}[t]
\centering
\includegraphics[scale=.38]{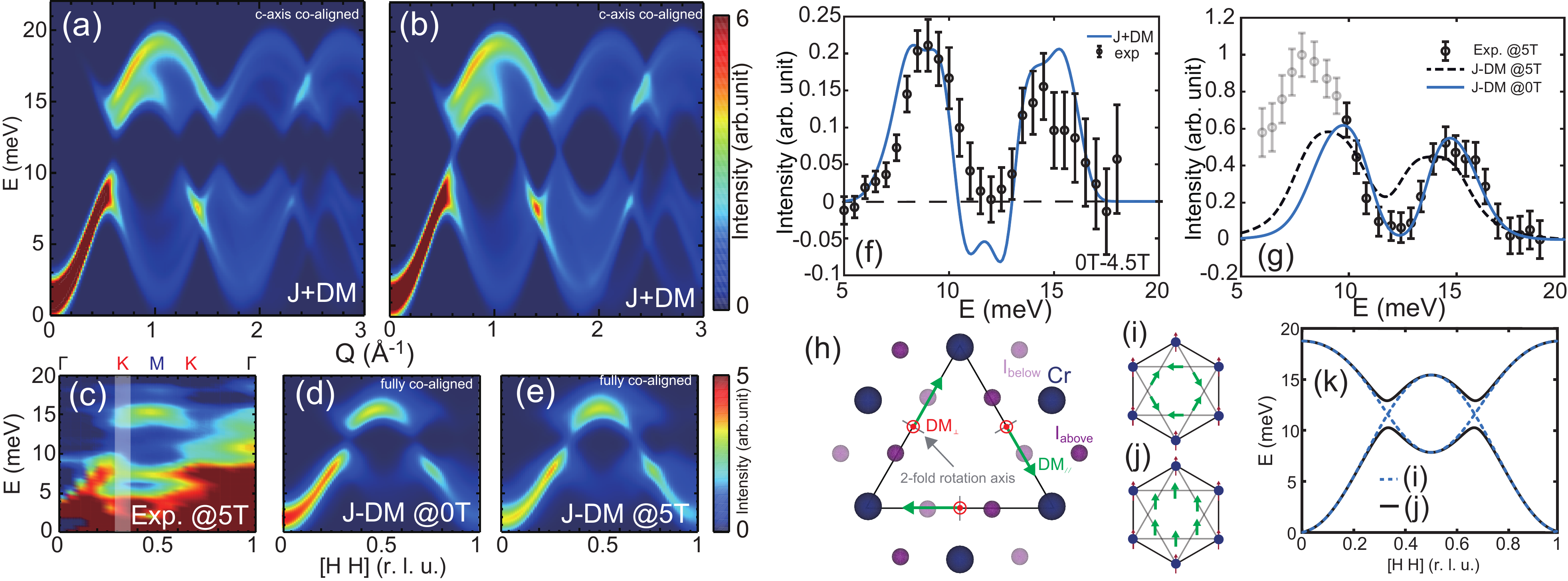}
\caption{\label{fig6}The magnetic field effect on the Heisenberg-DM model. 
(a, b)  Calculated $E$-\textbf{Q} spectra of $c$ axis aligned CrI$_3$ using the Heisenberg-DM Hamiltonian with 0 and 4.5 T in-plane fields, respectively.  (c) Spin waves of a fully co-aligned CrI$_3$ single crystals near the Dirac point along the [$H$, $H$] direction with a 5 T in-plane magnetic field. (d, e) Heisenberg-DM model simulation with 0 and 5 T in-plane field, respectively. (f) The effect of a magnetic field on spin wave dispersion near Dirac point and its comparison with the Heisenberg-DM calculations. (g) Constant-{\bf Q} cut at the Dirac point ($[H, H]= (0.3, 0.37)$) on the experimental data and Monte-Carlo simulations. The experimental data has a constant background subtracted. The gray dots show intensity increasing due to higher instrumental background. (h) Schematics of the in-plane DM interaction of a triangular sublattice in one Cr hexagon.  The in-plane component of the DM interaction (DM$_{||}$) is perpendicular to the 2-fold rotation axis between the two NN Cr ions according to the Moriya’s rule. (i) In-plane DM interactions respecting the 3-fold symmetry of the lattice. (j) An example of in-plane DM interactions breaking the 3-fold symmetry of the lattice. (k) The calculation of spin wave dispersion with in-plane spins and in-plane DM interactions shown in red dashed lines (i) and black solid lines (j). Here DM$_{||}= 0.17$ meV.
}
\end{figure*}

Figure 4(a) and 4(b) show expected spin wave spectra at $L=3$ and 4.5, respectively, calculated using a Heisenberg Hamiltonian 
with magnetic exchange parameters specified in Table I. Near the Dirac points, we see spin gap-like features at $K_1$ and $K_2$ due to shifted acoustic-optical spin wave touching points, and there is no true spin gap near the Dirac points. 
In addition, there is no evidence of Dirac nodal line winding in the spin wave 
spectra [Figs. 2(f,g)].
Figures 4(c) and 4(d) compare calculated and observed in-plane wave vector dependence of 
spin waves at the Dirac point with $L=3$.  Figures 4(e) and 4(f) are similar in-plane cuts with $L=4.5$.  By comparing cut through
the Dirac points with model calculations at $L=3, 4.5$ [Figs. 4(g,h)], 
we find that spin wave intensity at the Dirac energy peaks at the $K$ point and is independent of $L$, clearly different from the calculation. Therefore,
we conclude that the observed Dirac spin gap cannot 
arise from the electron correlation effects as discussed in Ref. \cite{Ke2021}.

Since both the NNN DM or Kitaev models can describe spin waves of CrI$_3$ \cite{LBChen2}, it will be important to determine which microscopic
model is correct.  One way to separate these two scenarios is to do an inelastic neutron scattering experiment on CrI$_3$ with a magnetic field applied within the $ab$-plane. The easy axis of spins in CrI$_3$ is parallel to the $c$ axis, but a magnetic field of 3 T will turn the spin to the $ab$-plane with almost zero out-of-plane component \cite{McGuire_CrI3_2014}. This change of the FM ordered moment direction will nullify the NNN DM term by making ${\bf A}_{ij}$ and ${\bf S}_i\times{\bf S}_j$ perpendicular to each other with vanishing $H_{DM}$, and therefore close the NNN DM interaction-induced spin gap at the Dirac points [Fig. 1(d)]. This is similar to the 2D kagome lattice ferromagnet Cu[1,3-
benzenedicarboxylate(bdc)] [Cu(1,3-bdc)], where 
an out-of-plane magnetic field applied to align the
in-plane FM ordered moments along the $c$ axis is
found to also induce a DM interaction-induced spin gap 
at the Dirac points \cite{Chisnell_kagome_2015,Chisnell2016}. In contrast, if the spin gap at the Dirac point is induced by the Kitaev exchange, its field dependence will be anisotropic and dependent on the relative angle of 
the polarized spin with respect to the in-plane lattice orientation [Figs. 5(a-f)].

To test this idea, we performed inelastic neutron scattering experiments under in-plane magnetic fields 
on HYSPEC \cite{HYSPEC} with incident neutron energy of $E_i = 27$ meV (Fig. 5) and on 
 ARCS \cite{ARCS} with $E_i = 23$ meV (Fig. 6).  Figure 5(a) shows the geometry of the
experimental setups, where the applied magnetic fields are vertical in the honeycomb lattice plane. For HYSPEC experiments, we used $c$ axis aligned single crystals ($\sim$1 g) [see inset of Fig. 5(h)] and applied a field of 4.5 T, 
which is larger than the in-plane 
saturation field of 3 T \cite{McGuire_CrI3_2014} and 
sufficient to completely polarize the moment
in the CrI$_3$ plane. As a function of increasing field,
the spin gap at the $\Gamma$ point ($\approx 0.4$ meV) \cite{LBChen2} initially decreases to overcome the $c$ axis aligned moment, but then increases due to the increasing Zeeman energy 
\cite{SI}. \Dai{These results are consistent with the field dependence of the gap 
from either single-ion spin anisotropy or 
off-diagonal $\Gamma$ term in Kitaev interaction \cite{SI}.}

Figures 5(g) and 5(i) show the spin wave ${\bf Q}$-$E$ 
spectra at zero and 4.5 T field, respectively. While the overall spin wave intensity 
decreases at 4.5 T due to rotation of the spin moment direction from the 
$c$ axis to the CrI$_3$ plane, the spin gap 
near the Dirac point, marked by the white vertical line in Figs. 5(g,i), shows no obvious change. 
In the J-K-$\Gamma$ model, the spin gap opens
at the Dirac points because the NN Kitaev exchange interactions alternate between two different anisotropic bond-dependent terms along the zigzag bonds \cite{Kitaev_ILee2020}. Since the Kitaev interaction Hamiltonian $H_K$ is inherently sensitive to the spin orientations,
spin wave spectra of a J-K-$\Gamma$ model will change drastically when 
the moment direction of the spins is rotated from the $c$ axis to the in-plane direction 
by an externally applied magnetic field [Figs. 5(b-f)].  
Whereas a DM interaction induced spin gap would close uniformly under an in-plane field to preserve the six-fold in-plane symmetry of the spin wave dispersion, 
the Kitaev interaction induced spin gaps will respond anisotropically depending on the relative angles between the wave vector and field direction. 
Furthermore, the field-induced changes in spin wave spectra will not be limited around the Dirac points in the J-K-$\Gamma$ model.

\Dai{Figures 5(h) and 5(j) show calculated spin wave ${\bf Q}$-$E$ spectra using the 
J-K-$\Gamma$ Hamiltonian with $c$ axis and in-plane moment, respectively.
We used the exchange parameters that reproduce the zero-field spectra identically with the Heisenberg-DM model shown in Fig. \ref{fig3}(b) \cite{LBChen2}. While the zero field calculation agrees well with the data, the 4.5 T spin wave spectra are 
clearly different from that of the calculation. 
} The data points in Fig. 5(k) and 5(l) show energy dependent
spin waves across the Dirac point at 0, and 4.5 T,
respectively. \Dai{The solid lines are spin wave 
calculations using the J-K-$\Gamma$ Hamiltonian with $c$ axis and in-plane moments, confirming
that the Heisenberg-Kitaev Hamiltonian clearly fails to describe the magnetic field effect on 
spin waves.}

\Dai{Figures 6(a) and 6(b) show calculated spin wave ${\bf Q}$-$E$ spectra using the 
Heisenberg-DM Hamiltonian with $c$ axis and in-plane moment, respectively. Compared with the 
J-K-$\Gamma$ Hamiltonian in Figs. 5(h) and 5(j), the 
Heisenberg-DM Hamiltonian obviously agrees much better with the experimental data in Figs. 5(g) and 5(i). }
Figure 6(c) shows the ${\bf Q}$-$E$ dependence of spin waves
near the Dirac point with an in-plane applied field of 5.0 T at 5 K,
obtained on 
co-aligned single crystals of CrI$_3$ on ARCS.
Figures 6(d) and 6(e) are the corresponding spin 
wave spectra calculated using the 
Heisenberg-DM Hamiltonian.
The data points in Figure 6(f) show the magnetic field
difference plot obtained from Figs. 5(k) and 5(l).  It is clear that 
the solid line calculated from the 
Heisenberg-DM Hamiltonian can approximately describe the data \Dai{but with small deviation near the Dirac point [Figs. 6(f,g)]}.

From the above discussions, we see that the J-K-$\Gamma$ Hamiltonian clearly 
cannot describe the observed 
magnetic field dependence of spin waves in CrI$_3$. 
While the simple NNN Heisenberg-DM Hamiltonian can describe the overall spectra 
and its magnetic field dependence, \Dai{it may have difficulty in describing the magnetic field dependence of the Dirac spin gap}. Since the loss of translational symmetry between the two Cr sublattice spins of an ideal honeycomb lattice can open a spin gap at the Dirac points,  
 it is important to determine other possible origins for the observed Dirac gap.

\section{Discussion}

In previous work \cite{CrI3_nature,XXZhang2020,JCenker2021,Pressure_CrI3,Pressure2,MFCrI3}, $A$-type AF order of CrI$_3$ was found to be associated with the monoclinic structural 
phase either near the surface of the bulk or in thin layer form (for example, the bilayer of CrI$_3$). However, it is unclear why the AF order in bilayer CrI$_3$ has monoclinic crystal structure, which appears in the paramagnetic
phase above $T_C$ of bulk CrI$_3$ \cite{CrI3_nature,XXZhang2020,JCenker2021}. Using the NN AF and NNN FM interlayer coupling in the rhombohedral FM phase (Figs. 1 and 2), we estimate that the interlayer stacking is still FM in the bilayer limit \cite{SI},
thus ruling out rhombohedral AF bilayer structure.
If we change the crystal structure to monoclinic but maintain the NN and NNN
$c$ axis coupling in bulk CrI$_3$, the magnetic bonding energies are higher than 
that of the rhombohedral lattice structure. 
From Raman scattering of bilayer CrI$_3$, the sum of the 
interlayer AF coupling in monoclinic
structure was found to be $\sim$0.11 meV \cite{JCenker2021}. Assuming that the NNN
magnetic exchange is negligible, we estimate that the NN magnetic exchange in
monoclinic bilayer is $J_{c1}=0.037$ meV \cite{SI}. Table II summarizes 
the total magnetic bonding energy for one Cr$^{3+}$ atom in different lattice 
and magnetic structures \cite{SI}. We find that the
FM bilayer rhombohedral structure should be more favorable than the AF bilayer monoclinic
structure, contrary to the observation. Since Raman experiments 
can only deduce total magnetic exchange along the $c$ axis, 
we are unable to determine the actual NN and NNN magnetic exchange couplings in the
monoclinic structure. Nevertheless, the observed AF order in the monoclinic bilayer suggests that such a phase has lower ground state energy compared with that of the FM rhombohedral structure in bulk or bilayer CrI$_3$.
As the hydrostatic pressure applied on the AF bilayer CrI$_3$ can
reduce the interlayer spacing and reintroduce the rhombohedral 
FM state \cite{Pressure2}, we expect that the monoclinic bilayer CrI$_3$ should have
 larger $c$ axis AF exchange and lattice parameter compared with that
of the rhombohedral bilayer. This is also consistent with a reduced $c$ axis lattice
constant below $T_C$ in bulk CrI$_3$ \cite{McGuire_CrI3_2014}, and recent
simulations of transport measurements suggesting that the layers may expand along 
the $c$ axis to minimize interaction energy and stabilize a different magnetic coupling \cite{ZWang2018,Prieto2011}.  We note that the collinear 
AF order in iron pnictides also expands the lattice parameter along the 
AF ordering direction \cite{Cruz2008,Zhao2008}. 
While the NNN interlayer exchange couplings of bulk CrI$_3$ ultimately determined its FM ground state, the AF interlayer coupling 
prevails in the monoclinic bilayer CrI$_3$  \cite{SI}. These results suggest that the monoclinic to 
rhombohedral structural phase transition in CrI$_3$ is driven by reducing the 
interlayer magnetic exchange energy.

\begin{table}[]
\centering
\begin{tabular}{c|c|c|c}
\hline
Structures & $J_{c1}$ (meV) & $J_{c2}$ (meV) & Energy (meV) \\
\hline
Rhom-bulk, FM & 0.048 & -0.071 & -1.33 \\
Rhom-bilayer, FM & 0.048 & -0.071 & -0.66 \\
Mono-bulk, FM & 0.048 & -0.071 & -0.21 \\
Mono-bilayer, FM & 0.048 & -0.071 & -0.10 \\
Mono-bulk, AF & 0.037 & 0 & -0.33 \\
Mono-bilayer, AF & 0.037 & 0 & -0.17 \\
\hline
\end{tabular}
\caption{The estimated magnetic bonding energies associated with each 
Cr$^{3+}$ atom in various crystal structure and exchange couplings \cite{SI}.
Rhom and Mono indicate rhombohederal and monoclinic lattice structures, respectively. 
In the hypothetical Mono-bulk and Mono-bilayer case, the NN and NNN magnetic
exchange couplings are asssumed to be the same 
 as those of Rhom-bulk and Rhom-bilayer, revealing that the FM rhombohederal 
lattice structure has lower magnetic bonding energy. 
}
\label{tab:my_label}
\end{table}

\Dai{Although our data ruled out pure Kitaev interaction and electron correlations as the microscopic origins of the observed Dirac spin gap,
there may be other interactions in addition to the NNN DM that contribute to the Dirac spin gap.}
We consider several possibilities. First, reducing the bulk structural symmetry from rhombohedral to monoclinic 
by itself will not open a spin gap at the Dirac point because 
such structural phase transition does not change the inversion symmetry of the 
Cr honeycomb sublattice. If additional structural 
deformations are present due to,  for instance,  thermal effects,  the inversion center between the first NNs would be removed.  This incidentally would allow 
DM interactions to exist at that level.  
Nevertheless,  we show in the Supplementary Section \cite{SI}  
that the inclusion of DM at the first NNs does not open a gap at the Dirac point. 
\Dai{We also consider a Heisenberg model with both the NNN DM and Kitaev interaction \cite{SI}. By using Heisenberg-DM Hamiltonian with different Kitaev interaction strength that fits spin wave spectra at 0 T, we can compare the expected and observed 
spin waves under 4.5T field and in-plane spin. The result indicates that the Kitaev term should be near zero in order to get the best fit to the 4.5 T spin wave spectra \cite{SI}.}

Alternatively,  magnon-magnon interactions may potentially 
affect $H_{DM}$ that can result in a gap at the Dirac point.  
When higher-order Holstein-Primakoff transformations 
are considered in the description of the spin interactions in CrI$_3$,  
3-operator products arise which may contribute to the gap \cite{SI}.  
However,  since magnon-magnon interactions in most magnetic 
materials are weakly energy and wave vector dependent,  and typically 
occur at energies above the single magnon scattering,  
they are unlikely to give rise to the observed spin gap at the Dirac points.

Finally, we envision two mechanisms that may allow the spin gap at the Dirac point to remain open under an in-plane spin polarizing field:
the first is the sublattice symmetry breaking; and the second is the three-fold rotational symmetry breaking of the ideal honeycomb lattice of
CrI$_3$.

We first discuss the possible sublattice symmetry breaking of an ideal honeycomb lattice.
From spin wave spectra in Figs. 2 and 3, we know that 
the two Cr$^{3+}$ ions of different sublattices within the honeycomb unit cell interact not only via the intralayer NN interaction $J_1$ but also the interlayer NN $J_{c1}$ which is AF and directly along the $c$ axis 
 [Fig. 1(a)]. Whereas both bonds are bisected by the structural inversion centers, respectively,  the interlayer AF exchange coupling $J_{c1}$ will favor a breaking of the inversion symmetry between the two Cr sublattice spins. As a result, if the two Cr$^{3+}$ ions within a unit cell have spins of unequal moments (due to environmental defects such as Cr and/or I vacancy) \cite{Meseguer-Sanchez}, an energy gap will appear at the Dirac points without significantly affecting spin waves at other wave vectors.

It is well known that the interlayer magnetic order in CrI$_3$ switches 
from AF to FM as the number of stacked vdW layers increase from 
bilayer to bulk, accompanied by a structural 
transition from monoclinic to rhombohedral stacking along 
the $c$ axis \cite{stacking1,stacking2,interlayer_SWJang,surfaceAFM_Niu,bilayer_Huang,bilayer_Sun}.  In addition, a small ($<3$ T) 
in-plane magnetic field can easily transform AF ordered
multilayer CrI$_3$ into a ferromagnet \cite{surfaceAFM_McCreary}. 
Even in the bulk samples, the surface layers are reported to 
have AF monoclinic structure that can be tuned by a $c$ axis 
aligned magnetic field of a few Tesla \cite{MFCrI3}.  
While these results indicate minor energy differences in 
rhombohedral and monoclinic structures of CrI$_3$,  
they suggest that the Cr honeycomb lattice may have subtle NN  
inversion symmetry breaking structural distortions 
that are responsible for the observed Dirac spin gap \cite{Meseguer-Sanchez}.

We next consider the field-induced breaking of the 3-fold symmetry of the in-plane DM vectors. Since the NNN DM interaction must involve the iodine atoms, the mirror symmetry of the simple honeycomb lattice is lost with only the two-fold rotation axis remaining \cite{SI}. As a result, 
the DM vector is not 
constrained to be out-of-plane and can have in-plane projections.  This argument holds
as long as the DM vector is perpendicular to the two-fold rotation axis
according to the Moriya's rule [Fig. 6(h)].  
In the case where no magnetic field is applied, the spins are aligned along the $c$ axis 
and only the DM vector component parallel to this direction can open the Dirac gap.  
In the situation where an in-plane applied magnetic field is strong enough to rotate the
$c$ axis aligned spins into the CrI$_3$ plane, the 3-fold symmetry of the in-plane DM vectors will cancel out when determining the spin wave energy at the $K$ point, thus yielding no contribution to the Dirac gap [Figs. 6(i,j)]. However, if the in-plane DM vectors breaking the 3-fold symmetry is induced by the applied field, then it will contribute to open a Dirac gap [Fig. 6(k)]. 
This will require a significant field-induced symmetry breaking of the in-plane DM whose energy scale should be similar to the out-of-plane DM terms ($\sim$0.17 meV). While a $c$ axis aligned magnetic field of a few Tesla 
is known to break the lattice symmetry of CrI$_3$ \cite{MFCrI3}, there is currently no direct experimental
proof that an in-plane magnetic field of a few Tesla would break the 3-fold symmetry of the crystalline lattice in CrI$_3$. Nevertheless, 
we could estimate a band gap of $\sim$2.0 meV using the parameters 
extracted from our data, in agreement 
with our experimental magnitude ($\sim$2.8 meV) \cite{SI}. 
This conjecture suggests that not only the Cr lattice contributes to the topological spin features observed in CrI$_3$ but also 
its halide sublattice.   

\section{Conclusions}

In summary, we used inelastic neutron scattering to study the impact of an in-plane magnetic field on spin waves of CrI$_3$.  At zero field, 
we completely determine the magnetic exchange couplings along the $c$ axis by carefully measuring $c$ axis spin wave dispersions at different in-plane wave vectors. We find
that the NN $c$ axis magnetic exchange coupling is AF and the NNN magnetic exchange couplings are FM. These results thus indicate 
coexisting AF and FM exchange interactions between the hexagonal layers of CrI$_3$. 
We also confirmed the presence of a spin gap at the Dirac points at zero field, and 
found that an in-plane magnetic field that 
can rotate the moment from $c$ axis to the CrI$_3$ plane \Dai{also modifies the spin wave spectra and spin gap at Dirac points}.
These results can conclusively rule out 
the J-K-$\Gamma$ Hamiltonian \Dai{and electron correlations as origins of the Dirac spin gap}.  While the field dependence of the Dirac spin gap
\Dai{may not be completely understood} within the NNN Heisenberg-DM Hamiltonian, 
the results suggest the presence of local sublattice or 3-fold rotational symmetry 
breaking of the ideal honeycomb lattice in CrI$_3$.  Our results therefore firmly establish 
the microscopic spin Hamiltonian in CrI$_3$, and 
provide a new understanding of topology-driven spin
excitations in 2D vdW magnets.

\section{Acknowledgments}
We are grateful to Franz G. Utermohlen, Nandini Trivedi, Adam Tsen, Zurab
Guguchia, Liuyan Zhao, and Rui He for helpful discussions. 
The neutron scattering and sample growth work at Rice is supported
by U.S. NSF-DMR-1700081 and by the Robert A. Welch
Foundation under Grant No. C-1839, respectively (P.D.).
The works of JHC was supported by 
the National Research Foundation (NRF) of Korea (Grant nos. 2020R1A5A1016518 and 2020K1A3A7A09077712).
EJGS acknowledges computational resources through the 
CIRRUS Tier-2 HPC Service (ec131 Cirrus\@EPCC Project) 
at EPCC (http://www.cirrus.ac.uk) funded 
by the University of Edinburgh and EPSRC (EP/P020267/1); 
ARCHER UK National Supercomputing Service (http://www.archer.ac.uk) via 
Project d429.  EJGS acknowledges the 
EPSRC Early Career Fellowship (EP/T021578/1) and 
the University of Edinburgh for funding support.  A portion of this research used
resources at the Spallation Neutron Source, 
a DOE Office of Science User Facilities operated
by the Oak Ridge National Laboratory.

{}

\end{document}